\documentclass[seceq]{jpsj2}
\makeatletter
\@addtoreset{equation}{section}
\makeatother
\newcommand{\eref}[1]{(\ref{#1})}
\usepackage{amsmath,amssymb,array}

\def\1{1\hspace*{-.3em}{\rm l}} 

\title{Additional Constants of Motion for a Discretization \\
of the Calogero--Moser Model}

\author{Hideaki \textsc{Ujino}$^{1}$\thanks{E-mail address: 
ujino@nat.gunma-ct.ac.jp}, 
Luc \textsc{Vinet}$^{2}$\thanks{E-mail address: 
luc.vinet@umontreal.ca}, 
Tetsu \textsc{Yajima}$^{3}$\thanks{E-mail address: 
yajimat@is.utsunomiya-u.ac.jp} and 
Haruo \textsc{Yoshida}$^{4}$\thanks{E-mail address: 
h.yoshida@nao.ac.jp}}

\inst{
$^{1}$Gunma National College of Technology, 
Maebashi, Gunma 371--8530\\
$^{2}$Universit\'e de Montreal, 
Montr\'eal, Qu\'ebec, Canada H3C 3J7\\
$^{3}$Department of Information Science, Faculty of Engineering, \\
Utsunomiya University, Utsunomiya, Tochigi 321--8585\\
$^{4}$National Astronomical Observatory of Japan, 
Mitaka, Tokyo 181--8588}

\abst{The maximal super-integrability of a
discretization of the Calogero--Moser model introduced by Nijhoff and Pang
is presented.
An explicit formula for the additional constants of motion is given.}

\kword{Calogero--Moser model,
super-integrability,
integrable discretization}

\begin{document}
\maketitle

\section{Introduction}
When one discretizes dynamical systems, it is hardly possible to avoid
modifying the original systems. Controlling such modifications is thus
a central problem in numerical analysis~\cite{Hairer}.
It would be ideal if a discretization 
conserves whole the structure of the original dynamical system
such as orbits in the phase space, 
constants of motion, integrability and so on. As an example of 
such ideal discretizations, a discretization of the Kepler problem,
which keeps all the constants of motion and the orbits in the phase space,
was discovered~\cite{Minesaki2002,Minesaki2004}. The Kepler problem is an
integrable system that has a set of mutually independent and 
Poisson commutative constants of motion, whose number is the same as the
degrees of freedom of the system. A dynamical system of $N$-degrees of freedom
which has mutually independent $2N-1$ constants of motion 
in the form of single-valued functions is called maximally
super-integrable and so is the Kepler problem. The above discretization
conserves super-integrability of the Kepler problem.

Among the family of one-dimensional integrable
systems with inverse-square interactions called
the Calogero--Moser--Sutherland models~\cite{vanDiejen_Vinet2000}, 
the Calogero model~\cite{Calogero1971},
which is the root of the family, the Calogero--Moser model~\cite{Moser1975}
of the rational and hyperbolic types are known to be maximally 
super-integrable\cite{Adler1977,Wojciechowski1983,Gonera1998}.
%
What we discuss here is the super-integrability of a discretization of 
the rational Calogero--Moser model, 
which is a classical dynamical system whose Hamiltonian is given by
\begin{equation}
  H:=\dfrac{1}{2}\sum_{i=1}^{N}p_i^2
  -\dfrac{1}{2}\sum_{\stackrel{\scriptstyle i,j=1}{i\neq j}}^{N}
  \dfrac{\gamma^2}{(x_i-x_j)^2},
  \label{eq:Calogero-Moser_Hamiltonian}
\end{equation} 
where $\gamma$, $N$, $p_i:=p_i(t)$ and $x_i:=x_i(t)$ are
the coupling parameter, the number of particles,
the momentum and the coordinate of the $i$-th particle at the time $t$,
respectively.

It will be no exaggeration to say that the Calogero--Moser model
represents the models of Calogero--Moser--Sutherland type
since the Lax formulation and a systematic
construction of the constants of motion for the model
was discovered earlier than those for any other models of
the family~\cite{Moser1975}.
Moser constructed a set of $N$ constants of motion
that are independent of each other.
Later, mutual Poisson commutativity of the constants of motion
of Moser-type was proved and  
the integrability of the model in Liouville's sense was 
thus established~\cite{Wojciechowski1977,Avan1993}. 
Furthermore, it turned out that the model had $N-1$ additional
constants of motion which are independent of the Moser-type ones and
independent of each other as well~\cite{Wojciechowski1983}.
This concludes the maximal super-integrability of the Calogero--Moser model.

A time-discretization of the Calogero--Moser model that conserves
the Moser-type constants of motion was presented by Nijhoff and 
Pang~\cite{Nijhoff-Pang1994}, which was reformulated into a more
convenient form by Suris~\cite{Suris2003}.
The aim of the paper is to show that
the maximal super-integrability of the Calogero--Moser model
holds good even after the above time-discretization: in other words,
the time-discretization of the Calogero--Moser model has $N-1$ additional
constants of motion, which are independent of the Moser-type ones 
and independent of each other at the same time. In \S\ref{sec:2}, 
we shall
give a brief summary of the discretization of the Calogero--Moser model and its
discrete Lax form. In \S\ref{sec:3}, we shall
explicitly construct $N-1$ additional constants of motion
of the discrete Calogero--Moser model. Concluding remarks are
summarized in \S\ref{sec:4}.

\section{The Discrete Calogero--Moser Model}\label{sec:2}
Throughout the paper, 
we employ Suris' formulation of the discrete Calogero--Moser model,
which is given by the following discrete symplectic map 
$(x_{i,n},p_{i,n})\rightarrow(x_{i,n+1},p_{i,n+1})$, 
$i=1,2,\ldots,N,$
\begin{equation}
  \begin{split}
    & 1-\Delta tc_0^{-1}p_{i,n}=\sum_{j=1}^N\dfrac{c_0}{x_{j,n+1}-x_{i,n}+c_0}
      -\sum_{\stackrel{\scriptstyle j=1}{j\neq i}}^N
      \dfrac{c_0}{x_{j,n}-x_{i,n}}, \\
    & 1-\Delta tc_0^{-1}p_{i,n+1}
      =\sum_{j=1}^N\dfrac{c_0}{x_{i,n+1}-x_{j,n}+c_0}
      -\sum_{\stackrel{\scriptstyle j=1}{j\neq i}}^N
      \dfrac{c_0}{x_{i,n+1}-x_{j,n+1}},
  \end{split}
  \label{eq:Suris}
\end{equation}
where $\Delta t$, $x_{i,n}:=x_i(n\Delta t)$ and $p_{i,n}:=p_i(n\Delta t)$ 
denote the discrete time-step, the coordinate and the momentum of 
the $i$-th particle at the $n$-th discrete
time $n\Delta t$~\cite{Suris2003}.
The constant $c_0$ is defined by $c_0^2:=-\gamma\Delta t$.
In terms of the Lax pair, which consists of two $N\times N$ matrices
below,
\begin{equation}
  \bigl(L_{n}\bigr)_{ij}=p_{i,n}\delta_{ij}
  +\dfrac{\gamma}{x_{i,n}-x_{j,n}}(1-\delta_{ij}), \quad
  \bigl(\mathcal{M}_{n}\bigr)_{ij}=\dfrac{c_0}{x_{i,n+1}-x_{j,n}+c_0},
  \label{eq:Lax_pair}
\end{equation}
the discrete symplectic map~\eref{eq:Suris} is expressed 
by the discrete Lax equation,
\begin{equation}
  L_{n+1}\mathcal{M}_n=\mathcal{M}_nL_{n},
  \label{eq:dLax_Suris1}
\end{equation}
which is equivalent to
\begin{equation}
  \label{eq:key1}
  L_{n+1}=\mathcal{M}_n L_{n} \mathcal{M}_n^{-1}.
\end{equation}
The companion matrix $\mathcal{M}_n$ thus plays a role of
the time-evolution operator of the Lax matrix $L_n$.
With the aid of the trace identity ${\rm Tr}AB={\rm Tr}BA$ where
$A$ and $B$ are arbitrary $N\times N$ matrices
as well as the discrete Lax equation~\eref{eq:dLax_Suris1}, 
one confirms that
the trace of the power of the Lax matrix $L_{n}$ satisfies
\begin{align*}
  {\rm Tr}\bigl(L_{n+1}\bigr)^m & = {\rm Tr}\Bigl(\mathcal{M}_nL_{n}
  \mathcal{M}_n^{-1}\Bigr)^m
  = {\rm Tr}\bigl(L_{n}\bigr)^m.
\end{align*}
Thus the discrete Calogero--Moser model~\eref{eq:Suris} as well conserves
the Moser-type quantities, which are exactly the same as the $N$ constants
of motion of Moser-type in the continuous time case~\cite{Moser1975}, 
\begin{equation}
  \label{eq:Moser}
  I^{(m)}_n:={\rm Tr}\bigl(L_n\bigr)^m,\quad m=1,2,\ldots,N.
\end{equation}
The Moser-type quantities~\eref{eq:Moser} are single-valued for
they are rational functions of $p_{i,n}$'s and $x_{i,n}$'s.
In order to confirm the mutual independence of the
Moser-type quantities, all one has to do is to check their
explicit forms when $\gamma=0$,
\begin{equation}
  I_n^{(m)}\Bigr|_{\gamma =0}=\sum_{i=1}^N\bigl(p_{i,n}\bigr)^m,
  \label{eq:sym1}
\end{equation}
which is nothing but the power sums of $p_{i,n}$'s that
are indeed independent of each other.
Note that the Hamiltonian \eref{eq:Calogero-Moser_Hamiltonian}
corresponds to the second constant of motion of Moser-type,
$H\Bigr|_{t=n\Delta t}=I^{(2)}_n/2$.

The companion matrix $\mathcal{M}_n$ of the Lax pair~\eref{eq:Lax_pair}
satisfies another Lax equation,
\begin{equation}
  D_{n+1}\mathcal{M}_n=\mathcal{M}_nD_n
  +\mathcal{M}_n\Delta t L_n
  \bigl(I-\Delta t c_0^{-1}L_n\bigr)^{-1},
  \label{eq:dLax_Suris2}
\end{equation}
where $I$ is the identity matrix and 
$D_n:={\rm diag}(x_{1,n},x_{2,n},\ldots,x_{N,n})$.
The above relation~\eref{eq:dLax_Suris2} was the crucial key to the
solution of the initial value problem of the discrete symplectic
map~\eref{eq:Suris}. In the next section,
we shall show how the relation \eref{eq:dLax_Suris2} works
in a systematic construction of
$N-1$ additional constants of motion of the discrete 
Calogero--Moser model~\eref{eq:Suris}.

\section{Additional Constants of Motion}\label{sec:3}
Our main purpose is to confirm that the $N-1$ quantities below
\begin{equation}
  \label{eq:Wojciechowski}
  \begin{split}
    K_n^{(m)}:=& 
    {\rm Tr}D_n\bigl(I-\Delta t c_0^{-1}L_n\bigr)
    \bigl(L_{n}\bigr)^{m-1}{\rm Tr}L_n\\
    &
     -{\rm Tr}\bigl(L_{n}\bigr)^m
    {\rm Tr}D_n\bigl(I-\Delta t c_0^{-1}L_n\bigr),\quad 
    m=2,3,\cdots,N, 
  \end{split}
\end{equation}
are conserved by the discrete time evolution of 
the discrete Calogero--Moser model~\eref{eq:Suris} and that
they are independent not only of the Moser-type quantities~\eref{eq:Moser}
but also of each other. Note that the case $m=1$ is omitted
in eq.~\eref{eq:Wojciechowski} because $K_n^{(1)}=0$.

The discrete symplectic map~\eref{eq:Suris} is equivalent to
the discrete Lax equations~\eref{eq:dLax_Suris1} and~\eref{eq:dLax_Suris2}.
From the discrete Lax equations~\eref{eq:dLax_Suris1} 
and~\eref{eq:dLax_Suris2}, one obtains
\begin{equation}
  \label{eq:dLax_Suris3}
  D_{n+1}\bigl(I-\Delta t c_0^{-1}L_{n+1}\bigr)\mathcal{M}_n
  =  \mathcal{M}_nD_n\bigl(I-\Delta t c_0^{-1}L_n\bigr)
     +\mathcal{M}_n\Delta t L_n,
\end{equation}
which is rewritten as
\begin{equation}
  \label{eq:key2}
  D_{n+1}\bigl(I-\Delta t c_0^{-1}L_{n+1}\bigr)
  =  \mathcal{M}_nD_n\bigl(I-\Delta t c_0^{-1}L_n\bigr)\mathcal{M}_n^{-1}
     +\mathcal{M}_n\Delta t L_n\mathcal{M}_n^{-1}.
\end{equation}
The relation~\eref{eq:key2} gives the time-evolution of the matrix
$D_n\bigl(I-\Delta t c_0^{-1}L_n\bigr)$.
Using eqs.~\eref{eq:key1} and \eref{eq:key2} as well as the trace identity,
one can perform the calculation below,
\begin{equation}
  \begin{split}
  K^{(m)}_{n+1} = &{\rm Tr}D_{n+1}\bigl(I-\Delta t c_0^{-1}L_{n+1}\bigr)
  \bigl(L_{n+1}\bigr)^{m-1}{\rm Tr}L_{n+1}\\
  & -{\rm Tr}\bigl(L_{n+1}\bigr)^m
  {\rm Tr}D_{n+1}\bigl(I-\Delta t c_0^{-1}L_{n+1}\bigr) \\
  = &{\rm Tr}\mathcal{M}_n\Bigl(D_n\bigl(I-\Delta t c_0^{-1}L_n\bigr)
  +\Delta tL_n\Bigr)\mathcal{M}_n^{-1}
  \bigl(\mathcal{M}_nL_{n}\mathcal{M}_n^{-1}\bigr)^{m-1}
  {\rm Tr}\mathcal{M}_nL_n\mathcal{M}_n^{-1}\\
  &-{\rm Tr}\bigl(\mathcal{M}_nL_{n}\mathcal{M}_n^{-1}\bigr)^m
  {\rm Tr}\mathcal{M}_n\Bigl(D_n\bigl(I-\Delta t c_0^{-1}L_n\bigr)
  +\Delta t L_n\Bigr)\mathcal{M}_n^{-1}\\
  =& {\rm Tr}D_n\bigl(I-\Delta t c_0^{-1}L_n\bigr)
  \bigl(L_{n}\bigr)^{m-1}{\rm Tr}L_n-{\rm Tr}\bigl(L_{n}\bigr)^m
  {\rm Tr}D_n\bigl(I-\Delta t c_0^{-1}L_n\bigr) \\
  &+\Delta t\Bigl({\rm Tr}\bigl(L_n\bigr)^m{\rm Tr}L_n
  - {\rm Tr}\bigl(L_n\bigr)^m{\rm Tr}L_n\Bigr) \\
  =& K_n^{(m)},
  \end{split}
  \label{eq:crucial}
\end{equation}
which proves the conservation of $K_n^{(m)}$. As one can observe
in the third line of eq.~\eref{eq:crucial}, cancellation of the
unwanted terms derived from the second term in the r.h.s.~of
eq.~\eref{eq:key2} is crucial.
The additional constants of motion~\eref{eq:Wojciechowski}, which
we call the Wojciechowski-type quantities, are rational functions
of $p_{i,n}$'s and $x_{i,n}$'s.

When the coupling parameter $\gamma$ and the time-step $\Delta t$ are zero,
the $N-1$ constants of motion $\{K_n^{(m)}\}$~\eref{eq:Wojciechowski} 
reduces to symmetric polynomials of $p_{i,n}$'s and $x_{i,n}$'s, 
\begin{equation}
  \lim_{\gamma\rightarrow 0}\lim_{\Delta t\rightarrow 0}K_n^{(m)}
  =\sum_{i=1}^N
  x_{i,n}\bigl(p_{i,n}\bigr)^{m-1}\sum_{j=1}^N p_{j,n}
  -\sum_{i=1}^N \bigl(p_{i,n}\bigr)^m\sum_{j=1}^N x_{j,n}.
  \label{eq:sym2}
\end{equation}
Though it is less trivial than the mutual independence of 
$I_n^{(m)}\Bigr|_{\gamma=0}$ \eref{eq:sym1},
the quantities~\eref{eq:sym2} are independent of those 
in eq.~\eref{eq:sym1} and independent of each other, too.
Its verification is essentially the same as that for the
additional constants of motion of Wojciechowski-type in the
continuous time case~\cite{Wojciechowski1983}. Thus we find
that the discrete symplectic map~\eref{eq:Suris} has 
$2N-1$ constants of motion $\{I_n^{(m)},K_n^{(m)}\}$, 
which are independent of each other and single-valued as well.
This concludes that the discrete
symplectic map~\eref{eq:Suris} gives not only an integrable, 
but a maximally super-integrable discretization 
of the Calogero--Moser model~\eref{eq:Calogero-Moser_Hamiltonian}.
This property of the discrete symplectic map~\eref{eq:Suris} corresponds to 
the maximal super-integrability of the Calogero--Moser model in the
continuous time case~\cite{Moser1975,Wojciechowski1983}.

\section{Concluding Remarks}\label{sec:4}
The main result of the paper is the construction of the $N-1$ additional
constants of motion~\eref{eq:Wojciechowski} besides the known $N$
constants of motion~\eref{eq:Moser} 
of the discrete symplectic map~\eref{eq:Suris}.
The result concludes the maximal
super-integrability of the discrete Calogero--Moser
model~\eref{eq:Suris}.

It should be remarked that the $N-1$ additional constants of motion
$\{K_n^{(m)}\}$ are not exactly the same as those for the Calogero--Moser
model in the continuous time case~\cite{Wojciechowski1983},
because of the additional term proportional to $\Delta t$ 
in their construction~\eref{eq:Wojciechowski}.
In the continuous time limit $\Delta t\rightarrow 0$, however, 
the additional constants of motion~\eref{eq:Wojciechowski} reduces
to exactly the same additional constants of motion 
for the non-discrete Calogero--Moser model discovered by
Wojciechowski~\cite{Wojciechowski1983}.
In other words, $K_n^{(m)}$ is a one-parameter deformation
of the additional constants of motion of Wojciechowski-type in the
continuous time theory.
Since the orbit in the $2N$-dimensional phase space of 
the maximally super-integrable model of $N$ degrees of freedom is 
uniquely determined by its $2N-1$ constants of motion,
the orbit of the discrete symplectic
map~\eref{eq:Suris} in the $2N$-dimensional phase space differs from
that of the Calogero--Moser model in the continuous time case, even
though both evolve from the same initial values.
The former gives a one-parameter deformation of the latter.

When one deals with the Calogero--Moser model, its pairwise interactions
are usually repulsive. The discrete symplectic 
map~\eref{eq:Suris} with a pure imaginary $\gamma$
conserves the Calogero--Moser 
Hamiltonian~\eref{eq:Calogero-Moser_Hamiltonian} with repulsive
interactions. In this case, however, its solution
becomes complex in general. Thus in the physical sense,
the discrete symplectic map cannot describe a discrete version
of the Calogero--Moser model with repulsive interactions.
On the other hand, another super-integrable discretization of the
Calogero--Moser model is given from the super-integrable discretization
of the Calogero--Moser model with an external harmonic 
confinement~\cite{UVY}.
This discretization conserves exactly the same constants of motion
of the Calogero--Moser model in the continuous time case and hence
reproduces exactly the same orbit in the phase space.
Repulsive interactions can be dealt with as well.
Details on the comparison of the two different discretizations
will be presented in a separate paper.

\section*{Acknowledgements} Most of the work was carried out during
the short stay of H.U.~at CRM, Universit\'e de Montr\'eal hosted by L.V. 
H.U.~is grateful to the warm hospitality of the institute.
This author is also supported by the Grant-in-Aid for Young
Scientists (B) (No.~17740259) from the Ministry of Education, 
Culture, Sports, Science and Technology of Japan.
The work of H.Y.~is partially supported by a Grant-in-Aid for
Scientific Research of JSPS, No.~18540226.

\end{document}